# Mosaic pattern formation in exfoliated graphene by mechanical deformation

Maria Giovanna Pastore Carbone[1], Anastasios C. Manikas[1,2], Ioanna Souli[1], Chistos Pavlou[1,2] and Costas Galiotis[1,2]

[1]Institute of Chemical Engineering Sciences, Foundation of Research and Technology-Hellas (FORTH/ICE-HT), Stadiou Street, Platani, Patras, 26504 Greece.

E-mail: c.galiotis@iceht.forth.gr, galiotis@chemeng.upatras.gr

[2]Department of Chemical Engineering, University of Patras, Patras, 26504 Greece

## Abstract

*Graphene is susceptible to morphological instabilities such as wrinkles and folds, which result from the imposition of thermo-mechanical stresses upon cooling from high temperatures and/ or under biaxial loading. A particular pattern encountered in CVD graphene is that of mosaic formation. Although it is understood that this pattern results from the severe biaxial compression upon cooling from high temperatures, it has not been possible to create such a complex pattern at room temperature by mechanical loading. Herein, we have managed by means of lateral wrinkling induced by tension and Euler buckling resulting from uniaxial compression upon unloading, to create such patterns in exfoliated graphene. We also show that these patterns can be used as channels for trapping or administering fluids at interstitial space between graphene and its support. This opens a whole dearth of new applications in the area of nano-fluidics but also in photo-electronics and sensor technologies.*

# Introduction

Graphene and 2D materials exhibit interesting morphologies which have been classified as wrinkles, ripples and/or crumples[1, 2, 3] depending on their physical dimensions, topology and order. Such corrugations are of paramount importance for the potential applications of graphene since they can modify the electronic structure and carrier transport, alter surface properties, create electron/hole puddle and induce pseudomagnetic field in bilayers[4]. The formation of wrinkles, ripples and crumples has been attributed to several physical causes, such as (a) the thermal vibration of the 2D lattice[5], (b) edge instabilities, defects/ dislocations[6], (c) the mismatch between the thermal expansion of the 2D material and its host substrate[7, 8, 9] (d) the evaporation/removal of solvent and its surface tension[4, 10], (e) the relaxation of pre-strained substrate and its surface potential[11, 12, 13]. Recently, it has been demonstrated that wrinkles in the form of delaminated folds in supported graphenes can be formed either by uniaxial compression or uniaxial tension beyond a certain critical load, depending on the mode of loading. In the first case, the wrinkling direction is perpendicular to the compression axis whereas in tension, wrinkles of the same pattern originate parallel to the loading direction due to Poisson's contraction exerted laterally by the underlying polymer substrate[14].

Herein we demonstrate that in-situ uniaxial tensile test combined with Atomic Force Microscopy can be employed to highlight the formation of complex morphologies in graphene flakes, such as the creation of mosaic patterns similar to those observed in CVD graphenes upon cooling from high temperatures (~1000° C).

# Results and Discussion

*Formation of mosaic morphology upon mechanical loading*

In the present study we used a 1 mm-thick sheet of poly-methylmethacrylate (PMMA) as substrate to simply support monolayer (1LG) and bilayer (2LG) graphene flakes that were prepared by mechanical cleavage from HOPG (High Order Pyrolitic Graphite) with the scotch tape method[15]. Appropriate flakes deposited directly on the PMMA were located using the optical microscope and the exact thickness of each flake was identified by the corresponding Raman line of the 2D spectra.

Optical images of the examined monolayer and bilayer graphene flake supported on PMMA are shown in Figure 1a. The dimensions of the approximately rectangular 2LG flake were ~35 μm (length) and ~60 μm (width), and those of the 1LG flake are 30 μm (length) and ~35 μm (width), which were large enough to ensure efficient load transfer at all strains[16] in both the loading and the transverse directions. At rest the flakes appear to be quite flat having only a small and smooth fluctuation of ±0.5 nm in the out-of-plane direction which corresponds to the surface roughness of the polymer substrate, in agreement with measurements conducted previously[14] (Figure 1c). The exact nature of the flakes was identified by the corresponding Raman spectra of the *2D* peak (Figure 1b). The position frequency of the *2D* peak at the unstressed state of the 2LG is 2604 cm$^{-1}$ and of the 1LG is 2598 cm$^{-1}$ which are close to the wavenumber at 0% strain for the excitation line of 785 nm, hence, the flakes have no residual stress.

A universal mechanical micro-tester (Deben, MT200) was adjusted under an AFM microscope (Dimension Icon Bruker), which allowed the visualization of the flake morphology under incremental uniaxial strain levels (in steps of 0.1%). Very importantly, a dual threaded leadscrew drives the jaws symmetrically in opposite directions, keeping the sample centred in the field of view during the tensile loading. The graphene flakes were subjected incrementally to tensile strain and the AFM images were captured at various levels of loading. AFM images were acquired with a Scanasyst-air probe of stiffness ~0.40 N/m and the measurements of the topography of the flakes were selected by using the operating Peak Force Nano-mechanical mode. The critical strain for the out-of-plane wrinkling and the wrinkling pattern development beyond a critical strain has been experimentally observed and quantified by using the proposed methodology. The experimental findings for the flakes under pure uniaxial tension are presented in Figure 2. For the sake of brevity, only images recorded at 0.40%, 0.90%, 1.50% and 2.00% are shown. As reported recently[14], the formation of clear localized wrinkles induced by tension parallel to the direction of the applied tension are noted at strain less than 1%, which is far lower than the predicted tensile strain to fracture of suspended graphene estimated at ~30%[17]. The phenomenon of lateral wrinkling observed here is typical of stretched elastic membranes that - at the critical tensile strain - buckle to accommodate the in-plane strain incompatibility generated via the Poisson effect[18]. For instance, for the case of graphene in air subjected to axial loading, the formation of lateral wrinkling has been recently observed by Raman spectroscopy through the undulation of the Raman frequencies indicating a small, non-uniform strain field corresponding to a sinusoidal wave with a wavelength of the order of 500 nm[19]. As for the case of a simply supported graphene flake subjected to uniaxial tensile loading, it is in fact loaded in compression in the lateral direction due to the Poisson's contraction of the polymer. A recent study of the group had already shown the formation of lateral wrinkles in graphene supported on PMMA and an analytical model based on continuum mechanics theory was developed to examine the dependence of the critical strain for lateral wrinkling upon the level of adhesion between graphene and the underlying substrate[14]. As shown in the AFM images of Figure 2, these vertical wrinkles are aligned and multiplied, and as the tensile strain increases their amplitude also increases. For instance, for the 2LG, the full width at half maximum (FWHM) at 0.4% is ~20 nm with an amplitude of ~1-2 nm, which grows up to 40 nm and 7-8 nm respectively, at higher strain (2.0%) (Figure 2a). It is interesting noting here that lateral wrinkling in supported monolayer graphene initiates at higher strain, in agreement with theoretical predictions[14], and that the height of the wrinkles increases from 1-2 nm to around 5 nm, while a very slight increase of FWHM is observed (Figure 2b). As noted earlier[14], this is a usual buckling failure meaning that, when the lateral wrinkles form, the upper layer sticks up and leave a hollowed region between graphene and the underlying substrate thus resulting in a local failure of the interface. Furthermore, upon mechanical deformation, the length of the lateral wrinkles seems to grow as adjacent wrinkles consolidate.

It is interesting to note that, upon unloading, the lateral wrinkles previously created upon tension are still present since they have been detached from the substrate and due to the interface slippage, the system does not fully recover. In addition, the formation of new wrinkles now perpendicular to the loading direction is observed (Figure 3a-c). The coexistence of lateral wrinkles originated by transverse strains (Poisson's effect) and longitudinal wrinkles originated by axial compression upon unloading result in the formation of a mosaic morphology which is depicted clearly in Figure 3a. It is worth adding that the mosaic morphologies are very similar for both monolayer and bilayer graphene, with wrinkles of comparable height but with different FWHM (which is higher for 2LG). Furthermore, another substantial difference is the lateral dimension of the islands delimited by the wrinkles, which is 1.37±0.39 μm for 2LG and 0.74±0.28 μm for 1LG.

The formed wrinkles are in fact isolated folds which have been detached from the underlying substrate and control tests reveal that do not come from any undulations of the surface of the substrate (Supplementary Figure 1). This is evident primarily in as-supported flakes under compression[20, 21] and has also been simulated by computational modelling[14, 21]. It constitutes a unique form of failure initiation which is the main characteristic of compression (lateral or longitudinal) of such systems. On the contrary flakes fully embedded in a matrix have been found to exhibit the expected Eulerian sinusoidal mode of compression failure[22] due to the presence of constraints for fold creation by the surrounding matrix.

The threshold of fold formation in the lateral direction seems to be accompanied by interfacial failure or slippage of the whole flake giving rise to the formation of folds in the longitudinal direction (Figure 4a) upon unloading as mentioned above. Here it seems that beyond a certain critical tensile strain $\varepsilon_p$, which is the maximum strain that can be transferred from substrate to graphene, the graphene flakes cannot be strained any further and gradually relax by slippage[23] (see Figure 4b-i). Hence, if the maximum deformation applied to the substrate is higher than this value ($\varepsilon_m > \varepsilon_p$), no further strain can be transmitted to graphene due to interfacial slippage. Upon unloading at a certain critical point, $\varepsilon_0$ ($\varepsilon_0 = \varepsilon_m - \varepsilon_p$) the graphene flake will be subjected to compressive strain, which can be of sufficient magnitude to induce longitudinal wrinkling, if exceeds the critical strain for wrinkling $\varepsilon_{cr}^{comp}$. In particular, it can thus be easily postulated (see Supplementary Discussion for details) that the applied strain level $\varepsilon_{onset}^{mosaic}$ at which mosaic originates is

$$\varepsilon_{onset}^{mosaic} = \varepsilon_m - (\varepsilon_p + |\varepsilon_{cr}^{comp}|) \qquad (1)$$

In light of the above equation, the minimum strain to be applied to the substrate in order to obtain mosaic morphology upon loading/unloading is evidently $\varepsilon_p + |\varepsilon_{cr}^{comp}|$. For monolayer graphene supported on PMMA, the critical strain for slippage has been found ~1.5% (Figure 4b) and, as stated before, $\varepsilon_{cr}^{comp}$ has been experimentally and theoretically found equal to ~-0.3%[21]; hence, it is possible to give an estimation of the minimum strain to apply to the substrate in order to create the mosaic morphology for the system at hand, which is approximately 1.8%. The proposed mechanism has been verified by independent mechanical experiments combined with Raman spectroscopy (Figure 4b-ii), which highlight also the gradual release of compressive strain after mosaic formation. Moreover, it is important to note here that the maximum deformation $\varepsilon_m$ that the graphene/polymer system can withstand is limited by the failure behaviour of the polymeric substrate (Supplementary Figures 2 and 3) and, also, by the poor interfacial adhesion in graphene/polymer systems in the absence of covalent bonding or additional non-covalent binding interactions such as π–π interactions or hydrogen bonding.

The mechanically self-assembled mosaic pattern formed here is very similar to the island-like morphology of CVD graphene grown on a copper substrate[16]. In that case the high thermal peripheral (biaxial) stresses induced due to the mismatch of the thermal expansion coefficients between copper and graphene result in a compressive biaxial strain of 2–3% in the graphene at room temperature. Such residual strain is far higher than the critical strain required for graphene wrinkling (buckling) thus giving rise to a network of large out-of-plane wrinkles or folds[8]. It is interesting to note that the peripheral equi-biaxial strain creates circular mosaic morphologies in contrast to the square patterns obtained here. In our case, the final lateral size in the mosaic morphology is essentially determined by the density of wrinkles developed during the loading and the unloading stages. The higher dimensions of the domains developed in the 2LG flake is compatible to a lower wrinkles density, compared to 1LG, in agreement with theoretical predictions[14]. These features are strongly dependent on levels of adhesion between graphene and the underlying polymer; hence, the use of different substrates can be thus exploited to tailor the final morphology of the mosaic.

*Mosaic morphology as self-assembled wrinkle networks for fluid administration*

In the mosaic morphology, when the newly formed wrinkles interact with the existing lateral wrinkles form different types of junctions. High resolution AFM images acquired at those points reveal that the most evident features are T-, X- and Y-junctions (Figure 3d), as also predicted by Zhang et al[24]. These typical features of the mosaic could be of paramount importance since it may broaden graphene application in the areas of electronics, photonics, nano-fluidics and bio-engineering[4]. For instance, the strong coupling between localized deformation and the electronic structure that is expected at the junctions could be exploited by strain engineering. Furthermore, constructive use of the self-assembled wrinkle networks could be harnessed as surface texturing agent to direct cell alignment and morphology in tissue engineering[25] or for the creation of nano-channels in nano-fluidic devices transporting liquids. In light of possible application, the stability of the wrinkles to cyclic loading, which is particularly important for the off/on controlled applications of the mosaic, has been also verified (Supplementary Figure 4). It is known that graphene is impermeable to most gases and liquids[26, 27] and thus the as-presented wrinkle networks obtained via mechanical deformation could be seen, in principle, as interconnected channels where fluid could flow or be stored at the nano-scale. In this regard, we have investigated herein the feasibility of mosaic morphology as described above for the development of mechanically self-assembled nano-fluidic devices. This has been carried out in combination with methods suitable for the encapsulation of liquids in the area between graphene and its support[28]. More specifically, after the formation of mosaic morphology via mechanical deformation, the supported graphene bilayer has been exposed to a high humidity environment (RH = 90%) for several days. The selection of bilayer mosaic for this feasibility study has been encouraged by recent results of a molecular dynamics study of pressure-driven water transport through graphene bilayers which suggests that, as the channel width increases, the water flow rate increases monotonically and also the increase of channel thickness leads to weaker water-graphene interaction and thus higher water molecule mobility[29]. The AFM images acquired immediately after the exposure reveal that triangular and polygonal blisters nucleate at most junctions (Figure 5), in agreement with theoretical models assessing the stability of wrinkle networks to interstitial pressure[24]. These blisters are likely induced by water molecules trapped in the interstitial space between supported graphene and the polymeric substrate, as also postulated earlier[28], and are connected through the network of wrinkles previously formed via mechanical deformation. It is interesting to note that control experiments reveal that, blisters shrink to their initial configuration of junctions as shown in Figure 5, after the removal of the system from the high humidity environment at RT. In light of these preliminary observations, the mosaic morphology of graphenes presented herein could be seen as a network of nano-reservoirs interconnected by nano-channels. Fluid release from blisters/nano-reservoirs could be triggered in different ways, depending on desired application; for instance, it has been shown that graphene blisters can be controlled by an external electric field[30]. Furthermore, it is interesting to note that the small dimensions of the channels can pose challenges to nano-fluidic devices transporting liquids, as recently demonstrated by Xie et al.[31]. In particular, it has been shown that if the graphene channel diameter is small enough, water diffusing into the interstitial space underneath can freeze at room temperature due to the strong geometric confinement and, if the space inside graphene wrinkles is large enough, water becomes liquid again and flows[28].

## **Methods**

*Sample preparation and flakes localization*

Graphene flakes were prepared by mechanical exfoliation of graphite using the scotch tape method and deposited directly on a PMMA/SU- 8 substrate. The SU-8 (SU-8 2000.5 Micro-Chem) is a photoresist polymer and was spin coated on the top surface of a PMMA film at ~2000 rpm creating a layer of ~300 nm thickness. The samples were characterized by means of AFM (see below) and Raman spectroscopy, by InVia (Renishaw) microspectrometer with 2400 grooves/mm grating for the 785 nm laser excitation; Raman spectra were measured at and the laser power was kept at 1.2 mW on the sample to avoid laser-induced local heating. A

100x objective with numerical aperture of 0.95 was used in all cases. Experimental data were fitted by using Lorentzian functions, one for the monolayer and four for the bilayer graphene.

*Combined uniaxial tensile test and Atomic Force Microscopy*

The tensile experiments were conducted using a micro-tensile tester (Deben MT 200) that could be adjusted under AFM for simultaneously applying strain and acquiring images under AFM. The increment step of strain was 0.1% for all cases. The AFM images were obtained using Dimension Icon microscope (Bruker) operating in Peak Force Tapping mode using ScanAsyst-Air probes (stiffness 0.2−0.8 N/m, frequency ~80 kHz). No data treatment apart from line subtraction (retrace) to remove the tilt has been performed. Wrinkle amplitudes and half wavelengths were measured from line profiles manually for each wrinkle. Medians of the thus obtained parameters were then used to characterize the wrinkles in each profile. Also, lateral size of the islands delimited by the wrinkles has been estimated by averaging the length of two crossed lines across the island measured manually from line profiles.

*Combined uniaxial tensile test and Raman Spectroscopy*

Tensile experiments have been performed using the same set-up under the Raman microscope (InVia Renishaw microspectrometer) by using the 785 nm laser excitation. At each deformation step, several Raman spectra were acquired in different location of the 1LG flake under investigation (near the geometric centre) and experimental data were fitted by using one Lorentzian function.

*High humidity exposure experiments*

High humidity exposure experiments were performed at 90% RH and at RT in a custom-made climate chamber[32].

**Data Availability**

The authors declare that the data supporting the findings of this study are available within the article and its supplementary information files or from the corresponding author upon reasonable request.

**Acknowledgements**

The authors acknowledge the financial support of the European Research Council (ERC Advanced Grant 2013) via project no. 321124, "Tailor Graphene". the research project "Graphene Core 2, GA: 696656 – Graphene-based disruptive technologies", which is implemented under the EU-Horizon 2020 Research & Innovation Actions (RIA) and is financially supported by EC financed parts of the Graphene Flagship and the Open FET project "Development of continuous two-dimensional defect-free materials by liquid-metal catalytic routes" no. 736299-LMCat which is implemented under the EU-Horizon 2020 Research Executive Agency (REA) and is financially supported by EC.

**Author contribution**

CG, MGPC and ACM designed the experiments; IS and ACM prepared and characterized the samples. IS and ACM performed Raman tensile experiments. MGPC, ACM and CP performed the simultaneous AFM-tensile experiments. Finally, CG, MGPC and ACM analysed the collected data and wrote the manuscript. CG supervised the entire project.

**Competing Interests**

The authors declare no competing interests.

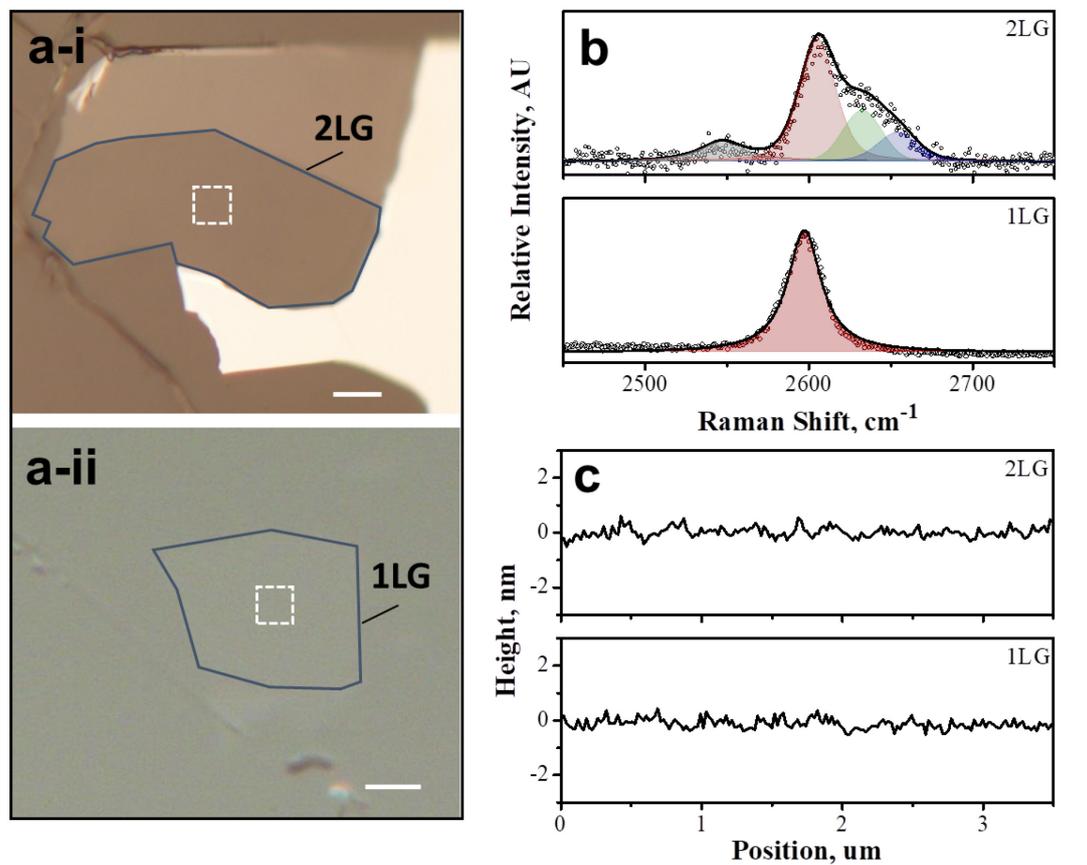

**Figure 1: Exfoliated graphene supported on PMMA substrate at rest.** a) Optical images of the examined 2LG and 1LG flakes delimited with a solid line to ease the visualization. The areas under investigation have been marked with dotted lines and the scale bar is 10 μm. b) Representative *2D* Raman spectra - acquired with a laser line of 785 nm - showing the typical spectral features of 2LG and 1LG graphene. Fits with Lorentzian functions (represented by the coloured peaks) are superimposed on the experimental data (dotted points). c) Typical height profiles of the flakes acquired across the dotted squares.

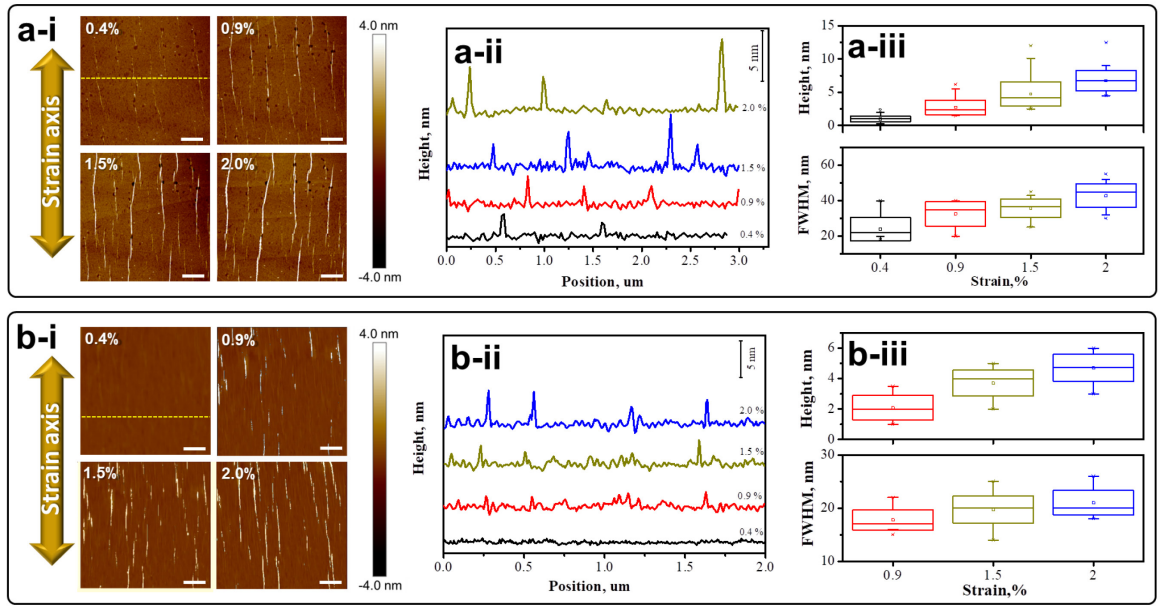

**Figure 2**: **Lateral wrinkling in supported bilayer (a) and monolayer (b) graphenes subjected to uniaxial loading.** (i) AFM images for the graphene flakes at different strain levels. The scale bar is 1 micron and the double arrow represents the direction of the applied tension. (ii) Height profiles representing the topography of the flakes for various levels of tensile strain at the cross-section line indicated in AFM image. The observed fluctuations of about ~0.5 nm between large wrinkles at all other strain levels correspond to surface roughness of polymer substrate. (iii) Statistical analysis of the height and the FWHM of the lateral wrinkles at different strain levels.

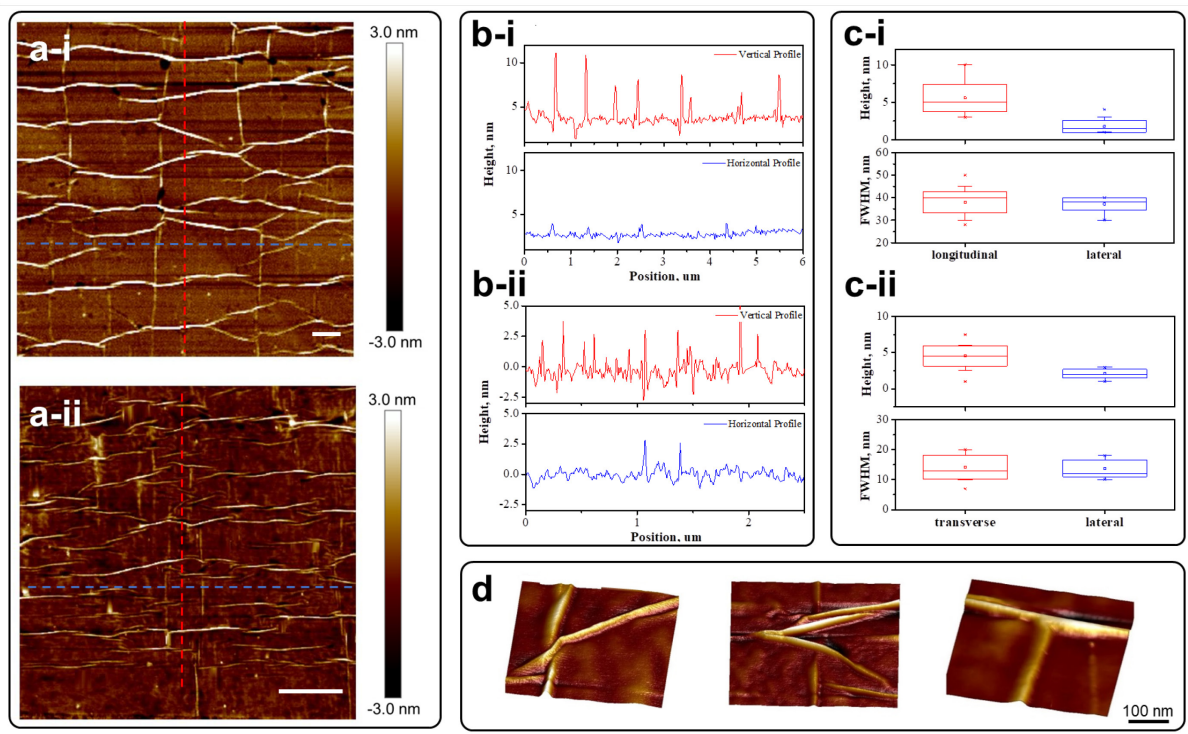

**Figure 3**. **Mosaic formation**. AFM images for the 2LG (a-i) and 1LG flakes (a-ii) after unloading (scale bar is 500 nm); vertical and horizontal height profiles representing the topography of the 2LG (b-i) and 1LG (b-ii) at the cross-section lines indicated in the AFM images; statistical analysis of the height and the FWHM of the longitudinal and lateral wrinkles in the mosaic originated in the 2LG (c-i) and 1LG flakes (c-ii). High-resolution images of X-, Y- and T-junctions originated - upon unloading - by the coexistence of lateral and longitudinal wrinkles (d).

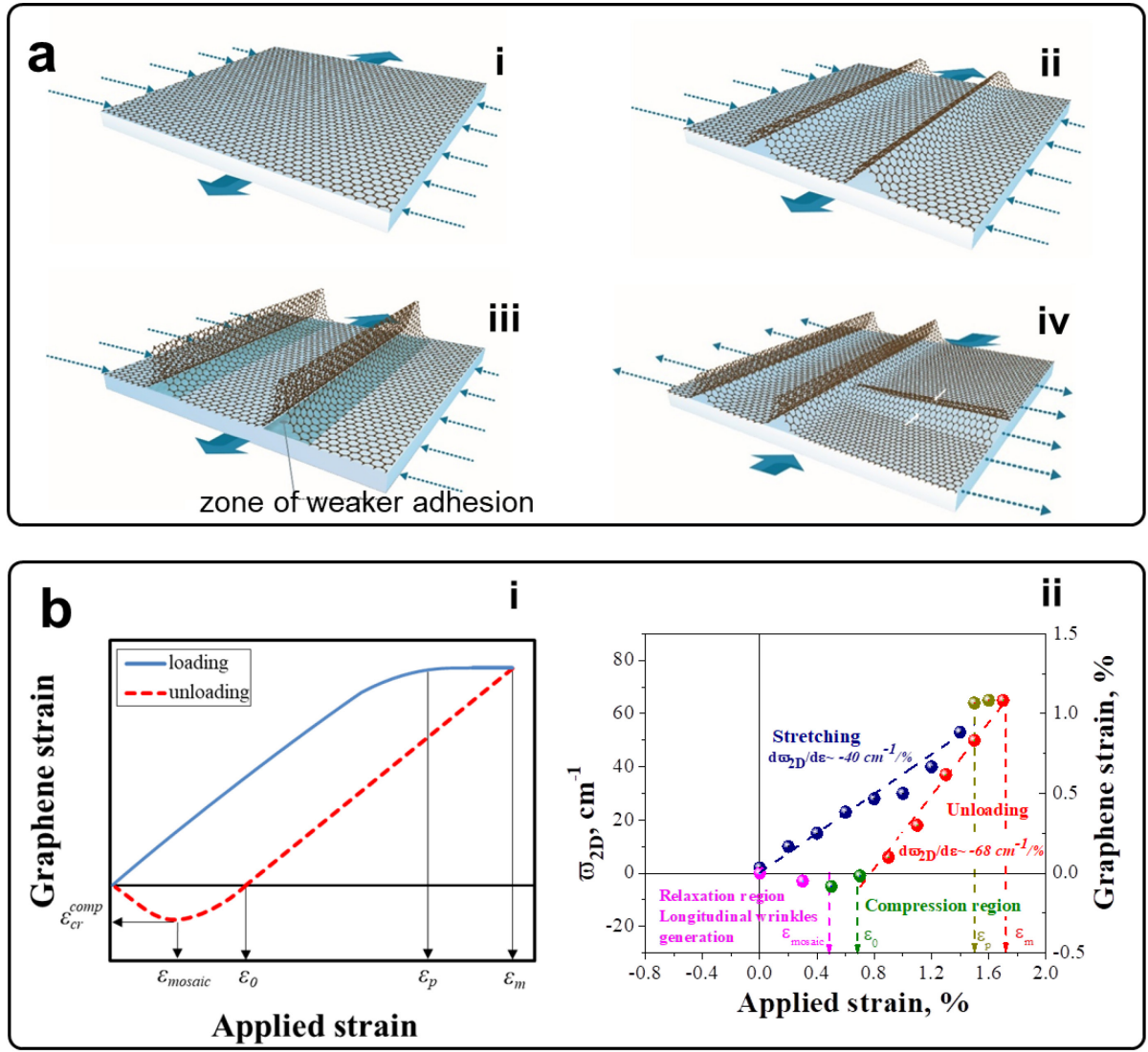

**Figure 4**: **Proposed mechanism of mosaic formation of a simply-supported graphene flake by mechanical deformation.** (a) Schematic of simply-supported graphene under uniaxial tension experiencing compression in the lateral direction due to Poisson's contraction of the substrate (indicated by the dotted arrows) (i). Upon loading – at the critical strain $\varepsilon_{cr}^{tens}$ - the lateral strain exerted by the Poisson's contraction of the underlying polymer induces graphene to buckle giving rise to lateral wrinkles (ii). As strain increases, the amplitude of these wrinkles increases. The wrinkles stick up above the PMMA substrate, leaving a hollowed region between graphene and the underlying substrate resulting in a local absence/lack of the interface (iii). Due to interface slippage, when the sample is subjected to unloading (iv), graphene tends to buckle in the perpendicular direction (as indicated by white arrows) at $\varepsilon_{onset}^{mosaic}$. The previously formed lateral wrinkles cannot be fully recovered and the intersection between the two families of wrinkles forms different junctions. The direction of the blue arrows indicates the loading/unloading of the substrate and graphene. (b) Representative loading-unloading cycle (i); 2D band Raman shift and graphene strain versus the applied strain for simply supported graphene on PMMA subjected to mechanical loading (ii)

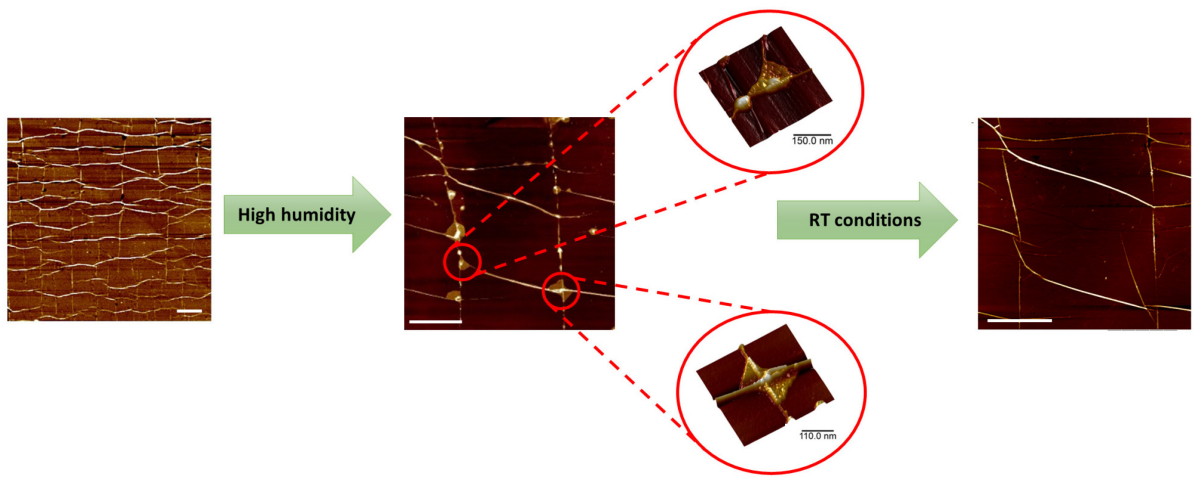

**Figure 5**: **A self-assembled wrinkle network for trapping or administering fluids.** AFM images concerning the formation and shrinkage of blisters in the mosaic morphology: after exposure at high humidity environment, topography of the bilayer and high resolution images reveal that typical polygonal and triangular blisters are induced by water molecules trapped in the interstitial space between supported graphene and the PMMA substrate; after removal from high humidity environment revealing the shrinkage of the blisters to initial junctions in a topographic detail of the bilayer. The scale bar is 1 micron.